\documentclass[%
 reprint,
showpacs,
 amsmath,amssymb,
 aps,
 prl,
 longbibliography,
 floatfix,
 superscriptaddress
]{revtex4-2}
\usepackage{xcolor}
\usepackage{soul}
\usepackage[utf8]{inputenc}	
\usepackage[T1]{fontenc}	
\usepackage[]{graphicx}		
\usepackage{xcolor}
\usepackage{gensymb}
\usepackage{amsmath, mathdots}
\usepackage{braket}
\usepackage{comment}
\usepackage[english]{babel}
\usepackage{color}

\begin{document}

\title{Parametric Amplification of a Quantum Pulse}

\author{Offek Tziperman}
\email{offekt@campus.technion.ac.il}
\affiliation{$\textup{Equal contributors.}$}
\address{Technion – Israel Institute of Technology, 32000 Haifa, Israel}
\author{Victor Rueskov Christiansen}
\email{victorrc@phys.au.dk}
\affiliation{$\textup{Equal contributors.}$}
\address{Department of Physics and Astronomy, Aarhus University, Ny Munkegade 120, DK-8000 Aarhus C, Denmark}
\author{Ido Kaminer}
\address{Technion – Israel Institute of Technology,  32000 Haifa, Israel}
\author{Klaus M\o lmer}
\email{klaus.molmer@nbi.ku.dk}
\address{Niels Bohr Institute, University of Copenhagen, Blegdamsvej 17, 2100 Copenhagen, Denmark}
\begin{abstract}
Creating and manipulating quantum states of light requires nonlinear interactions, but while nonlinear optics is inherently multi-mode, quantum optical analyses are often done with single-mode approximations. We present a multi-mode theory for the transformation of a quantum pulse by Hamiltonians that are quadratic in the field creation and annihilation operators. Our theory describes nonlinear processes, such as parametric amplification and squeezing, as well as all linear processes, such as dispersion and beam splitting. We show that a single input pulse feeds only two distinct output modes and, for certain quantum states, just one. Our theory provides the quantum states in the output modes, which are crucial for the application of pulses in quantum optics and quantum
information.
\end{abstract}
\maketitle

\noindent
\textit{Introduction}--- Quantum states of light are the key ingredients for some of the most promising quantum technologies \cite{photonic_quantum_tech}. Light pulses can propagate between stationary components in quantum networks \cite{internet_quantum}, and their applications include quantum key distribution \cite{QKD}, sensing beyond the standard quantum limit \cite{PhysRevD.26.1817, cavesnew, sensing}, bosonic error correcting codes \cite{GKP,GKP_exp}, and measurement based quantum computation \cite{measurment_based}. However, the preparation and manipulation of quantum states of light are more complicated for multi-mode traveling fields than for their stationary counterparts, such as the field in a single-mode cavity. This is because a continuum of frequency modes is available for the radiation \cite{RevModPhys.92.035005}.

The multi-mode nature of quantum light is of critical importance in any system involving nonlinearities, as in this case, the transformation of a quantum state will generally lead to a population of output pulse shapes correlated with the photon number state content [9, 10]. The output field is thus inherently multi-mode in nature, and the quantum properties of the state may not be accessible to the desired quantum information processing task. 

In this article, we deal with operations on traveling light pulses, governed by Hamiltonians that are of second order in the creation and annihilation operators. This class of Hamiltonians can describe beam splitters and interferometers, dispersion, diffraction and polarization rotation, but also parametric amplification, parametric down conversion and frequency conversion.

We are particularly interested in parametric amplification which is in the single (cavity) mode case described by the 
Hamiltonian ($\hbar = 1$), $H=\frac{i\xi}{2}((a^\dagger)^2-a^2)$, and the time evolution:
\begin{align}\label{single_mode_Bogolyubov}
    a(t) = \cosh(\xi t)a(0) + \sinh(\xi t)a^{\dagger}(0).
\end{align}      
This transformation is also referred to as squeezing, as it reduces the value (and uncertainty) of one of the field quadratures while the other quadrature is correspondingly amplified \cite{PhysRevD.23.1693}.
Both the squeezing and amplification properties are widely used in quantum optics for quantum sensing \cite{sensing}, quantum state tomography \cite{Kalash:23, Li:17} and for signal amplification and read-out of superconducting qubits \cite{TWPA_science_2015,TWPA_2023} or nanophotonics \cite{doi:10.1126/science.abo6213}. In its single mode version, parametric amplification has been proposed as part of the gate set for continuous variable quantum computing \cite{PhysRevLett.82.1784} and for creation and manipulation of Gottesman-Kitaev-Preskill (GKP) states for error correction \cite{PhysRevA.97.022341,GKP_OPA_singlemode_approx}. As there are now proposals and attempts to employ these schemes with traveling wave packet modes, it is pertinent to assess the multi-mode performance of the squeezing devices and ask to what extent Eq. \eqref{single_mode_Bogolyubov} applies to the transformation of the quantum state of a traveling pulse of radiation.  
\begin{figure}[h]
    \centering
    \includegraphics[width=\columnwidth]{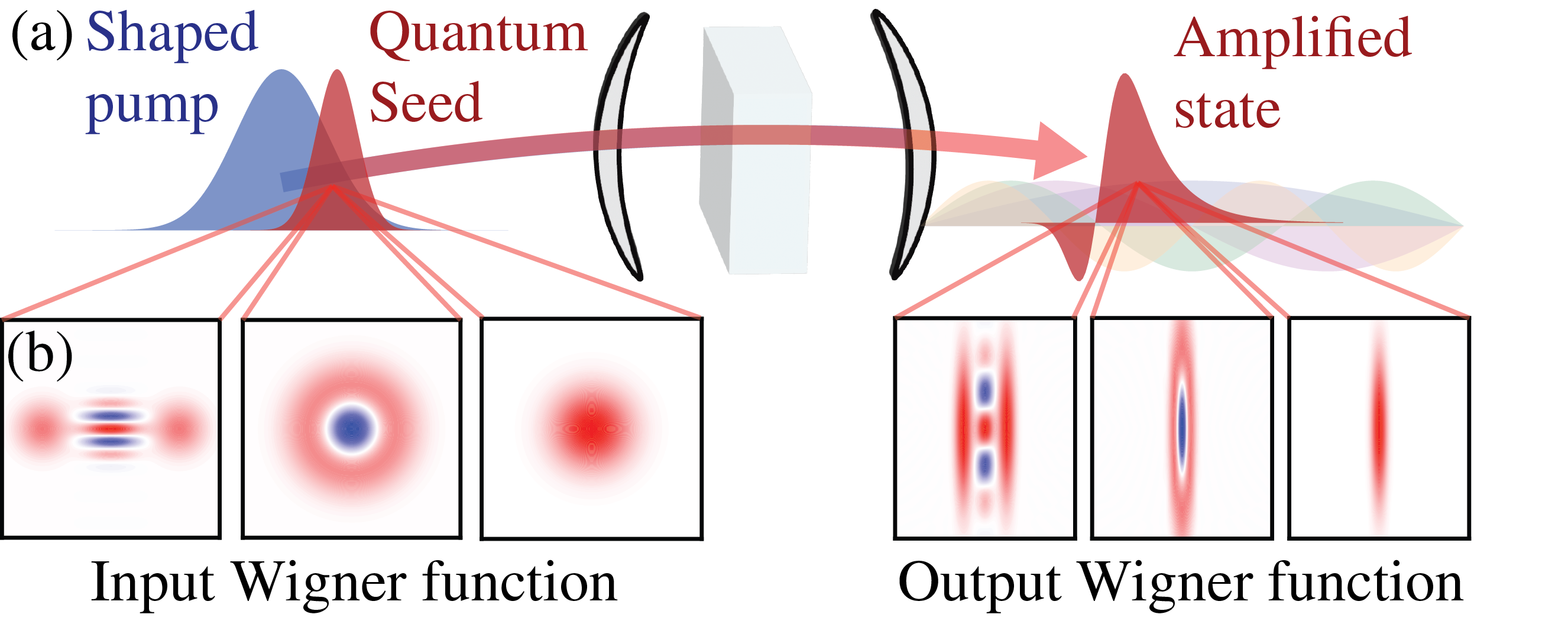}
    \caption{\textbf{Parametric amplification of quantum pulses}. \textbf{(a)} A non-linear medium in a cavity is driven by an arbitrarily shaped classical pump (blue) such that a single quantum pulse (red), transmitted through the cavity, is subject to parametric amplification. The cavity output field populates several spatio-temporal modes with components of the amplified pulse and with squeezed vacuum contributions. \textbf{(b)} When the input pulse is prepared in a  Schrödinger cat state, a single photon state, and a vacuum state (shown in the left panels), the mostly occupied output mode  is in an approximate, amplified (squeezed) version of these input states (shown in the right panels).}
    \label{setup}
\end{figure}

Here, we answer this question by presenting a general theory for the time evolution of quantum pulses of radiation subject to optical components that actuate quadratic Hamiltonians, such as dispersive and parametrically amplifying elements. Due to the multi-mode character of the problem, the transformation of quantum pulses by quadratic Hamiltonians is fundamentally different from situations where a cavity restricts the dynamics to a single mode. Still, even when subject to general, time-dependent Hamiltonians that are quadratic in creation and annihilation operators, a single-mode input quantum state transforms into an output retained in at most two spatiotemporal modes. We can thus calculate the state occupying these, or any other desired output wave packet mode, in which we recover a squeezed version of part of the initial state mixed with various amounts of squeezed vacuum. 


Figure \ref{setup} shows examples of parametric amplification of incident wave packet states. The lower panels in Figure \ref{setup} show Wigner functions of the input states and the states of the most populated output mode. While our theory takes the multi-mode character of traveling fields explicitly into account, it shows that it is still possible to obtain an approximate squeezing transformation from a single input to a single output mode.

\textit{Quantum state transformation by a parametric amplifier---}Our main goal is to determine the output of the amplifier given any arbitrary quantum state occupying an input wave packet $u(\omega)$. We treat here the case of guided propagation of a single transversal field mode, while further spatial and   polarization degrees of freedom are readily included, see Sec. III (SMIII)) in \cite{supp}.
We define: 
$a_{u,\textup{in}}^\dagger = \int_{-\infty}^\infty  u(\omega) a^\dagger(\omega)d\omega,$
where $a(\omega)$ are bosonic operators with the commutation relation $\left[a(\omega),a^{\dagger}(\omega')\right]=\delta(\omega-\omega')$. 

We consider a general quadratic Hamiltonian in the creation and annihilation operators  \cite{Christ_2013}:
\begin{align}\label{hamiltonian}
\begin{split}
    H = &\int \int d\omega d\omega'K(\omega,\omega')a^\dagger(\omega) a(\omega')\\+  &\int \int d\omega d\omega' J(\omega, \omega') a^\dagger(\omega) a^\dagger(\omega')  + \text{h.c.},
\end{split}
\end{align}
where $K(\omega,\omega')$ alone leads to a transformation of the input wave packet, while $J(\omega, \omega')$ is the amplitude for simultaneously creating a pair of photons in the output field at frequency $\omega$ and $\omega'$, and h.c. denotes the Hermitian conjugate.

The equations of motion can be solved to give the transformation of the operators in the Heisenberg picture \cite{Bogolyubov:1958se}:
\begin{align}\label{eq:in_out}
\begin{split}
    a_\textup{out}(\omega) = &\int d\omega' F(\omega, \omega') a_\textup{in}(\omega')\\ +  &\int d\omega' G^\ast(\omega, \omega') a_\textup{in}^\dagger(\omega'),
\end{split}
\end{align}
where $F(\omega, \omega')$ and $G(\omega, \omega')$ are uniquely determined by the functions $J(\omega, \omega')$ and $K(\omega,\omega')$ \cite{Christ_2013}, and we shall give examples in later sections of the article. 

The ideal transformation for parametric amplification of a single mode quantum pulse would amount to a single output mode squeezed as in Eq. \eqref{single_mode_Bogolyubov}. However, since the amplification applies to a continuum of field modes and is frequency-dependent, the output field may occupy mode functions that will generally differ from the input mode. We therefore move our focus to the calculation of the quantum state content of any given output mode with annihilation operator $a_{v,\textup{out}}=\int_{-\infty}^\infty v^\ast(\omega')a_\textup{out}(\omega')d\omega'$. First, we shall analyse the mode content of the output to find the optimal candidate $v$ function containing the transformed input quantum state, and in the following section, we shall calculate its actual quantum state content.

\textit{Modes at the output of the amplifier--- }To characterize the output field, we evaluate and  decompose the first order coherence function $g_1(\omega_1, \omega_2) = \braket{a^\dagger(\omega_1)a(\omega_2)}$ \cite{RevModPhys.92.035005}. Consider the situation where the input occupies a single wave packet,  $u(\omega)$, with a quantum state given by a density matrix $\rho_u$, and vacuum in all modes orthogonal to $u(\omega)$. We can then readily replace $\braket{a^{\dagger}_\textup{in}(\omega') a_\textup{in}(\omega'')}$  by $\braket{a^{\dagger}_u a_u}u^* (\omega')u(\omega'')$ and apply similar expressions for the other field-field correlation functions. With Eq. \eqref{eq:in_out} we find
\begin{widetext}
\begin{align}\label{eq:g1_fin1}
\begin{split}
g_{1}(\omega_{1},\omega_{2})&=\braket{a_{u}^{\dagger}a_{u}} \int d\omega'F^{*}(\omega_{1},\omega')u^{*}(\omega')\int d\omega''F(\omega_{2},\omega'')u(\omega'')+\braket{ a_{u}^{\dagger}a_{u}^{\dagger}} \int d\omega' F^{*}(\omega_{1},\omega')u^{*}(\omega') \int d\omega'' G^{*}(\omega_{2},\omega'')u^{*}(\omega'') \\ &+\braket{a_{u}a_{u}} \int d\omega'G(\omega_{1},\omega')u(\omega')\int F(\omega_{2},\omega'')u(\omega'')d\omega''+\braket{a_{u}^{\dagger}a_{u}} \int d\omega'G(\omega_{1},\omega')u(\omega')\int d\omega''G^{*}(\omega_{2},\omega'')u^{*}(\omega'')\\&+\int d\omega'G(\omega_{1},\omega')G^{*}(\omega_{2},\omega').
\end{split}
\end{align}
\end{widetext}
We observe that the coherence function is governed by both the input pulse's shape and quantum state. The first four terms contain correlation functions of the input field (for example $\braket{a_u a_u}$) while the last term is the squeezed vacuum output of the amplifier in the absence of any input pulse. 

By inspection of Eq. \eqref{eq:g1_fin1}, it turns out that the input quantum pulse gives rise to population of only two output modes, while a potentially infinite number of modes are populated by squeezed vacuum, 
\begin{align}\label{eq:g1_fin}
\begin{split}
g_{1}(\omega_{1},\omega_{2})=&
\underbrace{\sum^{2}_{i=1} n_i v_i^\ast (\omega_1) v_i(\omega_2)}_{\textup{dependent on input quantum state}} + \\ &\underbrace{\sum_{i=1}^\infty m_i w_i^\ast (\omega_1) w_i(\omega_2)}_{\textup{independent of input quantum state}}.
\end{split}\end{align}
This is demonstrated in detail in Sec. SMI \cite{supp}.   

In the special case of a coherent state input ($\rho_u=\ket{\alpha}\bra{\alpha}$) the operator expectation values factor, and the above expansion leads to only a single output mode seeded by the input field. For a Fock state input ($\rho_u=\ket{n}\bra{n}$) the output field occupies two modes, seeded by the input pulse. In fact, the condition for which the output occupies only a single mode is \cite{supp}
\begin{align} \label{eq:single_mode_condition}
    \braket{a_u^\dagger a_u} = |\braket{a_u a_u}|.
\end{align}
We note that these findings have intriguing and practical consequences for quantum state tomography using parametric amplification \cite{Kalash:23,doi:10.1126/science.abo6213}, where both input fed output modes must be considered, as well as the noise input from squeezed vacuum modes.

\textit{The quantum state in an output mode---} While the operation of the amplifier is entirely described by the multi-mode Bogoliubov transformation 
\eqref{eq:in_out}, this does not provide an immediate description of the transformation of a given quantum state input. We are faced, in fact, with an instance of the boson sampling problem \cite{10.1145/1993636.1993682}, and the number state representation of the output state is an unwieldy expression in terms of matrix permanents and the density matrix of the input state.     
The problem simplifies, however, when we restrict our interest to the quantum state of the output of the amplifier in any single  mode $v(\omega)$. Our theory applies to any output mode of interest, but a good choice for $v(\omega)$ is the most occupied single mode function at the output, say, $v_1(\omega)$ in Eq.\eqref{eq:g1_fin}. Assuming any output mode function $v(\omega)$, and by applying \eqref{eq:in_out}, we can formally write
\begin{align}
a_{v,\textup{out}}=\int v^{\ast}(\omega)a_{\textup{out}}(\omega)d\omega = \zeta a_{f,\textup{in}}+\xi a_{g,\textup{in}}^{\dagger}
\end{align}
where we have defined two new input mode functions through $f^*(\omega) =\int v^*(\omega') F(\omega', \omega) d\omega'/\zeta$ and $g(\omega) =\int v^*(\omega') G^*(\omega', \omega) d\omega'/\xi$, and $\zeta, \xi$ ensure their normalization (we omit the subscript $\cdot_\textup{in}$ in the following).

Notice a few key properties of this transformation: First, in the purely dispersive case, $\xi=0,\;\zeta=1$ and as a result, the annihilation operator of the output mode function $v(\omega)$ represents the exact same quantum state content as the one that occupied the input mode function $f(\omega)$ (which is not necessarily $u(\omega)$). This reflects that the quantum states are unchanged while traveling through a linearly dispersive element, but their mode functions change shape.

Second, in the amplifier case, where $\xi\neq0$, the functions $f,g$ are normalized but generally not orthogonal, and their relationship with the populated input mode $u(\omega)$ is not yet specified. To find the output quantum state in the mode $v(\omega)$, we must decompose the transformation into one that refers specifically to the input mode $u(\omega)$ and vacuum modes orthogonal to $u(\omega)$. Defining $\braket{f,g} =\int f^{\ast}(\omega)g(\omega)d\omega$, we decompose the modes into parallel and orthogonal components of the input mode $u$,
\begin{align}
\label{eq:threemodes}
\begin{split}
    a_{v,\textup{out}} &= \zeta \braket{f, u} a_u + \xi \braket{u, g} a_u^\dagger \\&+ \zeta\sqrt{1-|\braket{f,u}|^2}\braket{h,k} a_k + \xi \sqrt{1 - |\braket{u, g}|^2}a_k^\dagger \\
    &+\zeta \sqrt{1-|\braket{f,u}|^2}\sqrt{1-|\braket{k,h}|^2} a_s.
\end{split}
\end{align}
Explicit expressions for the mode functions for $k$, $h$, and $s$ are given in Sec. SMII in \cite{supp}.

Eq. \eqref{eq:threemodes} is a main result of our analysis. It shows how a single output mode captures a squeezed version of the potentially interesting state occupying the input pulse (first line). However, it is mixed with a squeezed vacuum component (second line) and a vacuum component (third line). 
In the absence of the terms in the second and third line, we recover unitary single-mode squeezing as in Eq. \eqref{single_mode_Bogolyubov}, while the presence of these terms in Eq. \eqref{eq:threemodes} will in general contribute added quantum noise. 

The transformation is effectively described by the Bloch-Messiah reduction \cite{PhysRevA.71.055801,spec_Bloch_messiah}, which states that a general multi-mode Bogoliubov transformation can be separated in 3 steps: first, a linear beam splitter transformation among the modes, then a sequence of single mode squeezing operations, and finally another beam splitter transformation. We thus find $\rho_{v\textup{,out}}=\textup{Tr}_{k,s}(U\rho_uU^{\dagger})$ with 
\begin{align}
U=U_{u,s}(\theta_3,\phi_3) U_{u,k}(\theta_2,\phi_2) S_u(r_1) S_k(r_2) U_{u,k}(\theta_1,\phi_1),
\end{align}
where the beam-splitter and squeezing transformations and parameters are specified in Sec. SMII in \cite{supp}. Note that these transformations act in a particularly simple manner on the states in the Wigner function representation, where they amount to linear transformations on the field quadrature variables \cite{Ekert}, given directly by the coefficients in Eq. \eqref{eq:threemodes}. By a straightforward extension of this procedure, one can also find the joint quantum state of the pair of output modes $v_1,v_2$ fed by the input mode, and, e.g., study their mutual entanglement properties, see Sec. SMII in \cite{supp} and the code repository\footnote{Link to the repository with Python code to find the output quantum state for a given transformation https://github.com/offektziperman/Amplifying-a-quantum-pulse}.
In the following, we shall restrict the analysis to the most occupied single mode $v_1$ for $v_\textup{out}$, and leave the study of other candidate output modes and the two-mode output for later investigation. 
     
\textit{Amplification by an Optical Parametric Oscillator---}
To demonstrate applications of our theory, we consider as a first example a degenerate Optical Parametric Oscillator (OPO) cavity with the cavity Hamiltonian
\begin{align} \label{eq:OPO-Hamiltonian}
    H_\textup{sys} = \Delta a^\dagger a +  \frac{i\xi(t)}{2}\left[ a^{\dagger 2} - a^2\right],
\end{align}
where $\Delta$ is the detuning of the cavity and $\xi(t)$ is the time-dependent parametric gain due to a pulsed drive. We assume an input quantum state occupying a single mode of temporal shape $u(t)$, coupled with the coherent amplitude $\sqrt{\gamma}$ to the cavity mode (the cavity loss rate is $\gamma$). Under the assumption of a Markovian coupling, we can relate the output field to the cavity field at time $t$ via the input-output relation \footnote{This input-output relation assumes unidirectional propagation of the field which may be ensured by a ring cavity, a circulator, or the phase matching in the OPA gain material in the absence of a cavity.} $a_\textup{out}(t)=a_\textup{in}(t)+\sqrt{\gamma}a(t)$ \cite{Input-Output-Theory}, and solving the Heisenberg equation of motion for the intra-cavity field, we obtain the explicit Bogoliubov transformation of the field operators in time domain in the SMV \cite{supp}. This yields directly the time domain correlation function $g_1(t_1,t_2)$, and the minimal basis of temporal modes, that efficiently describe the parametrically amplified input quantum state. 
\begin{figure}
    \centering
    \includegraphics[width=\columnwidth]{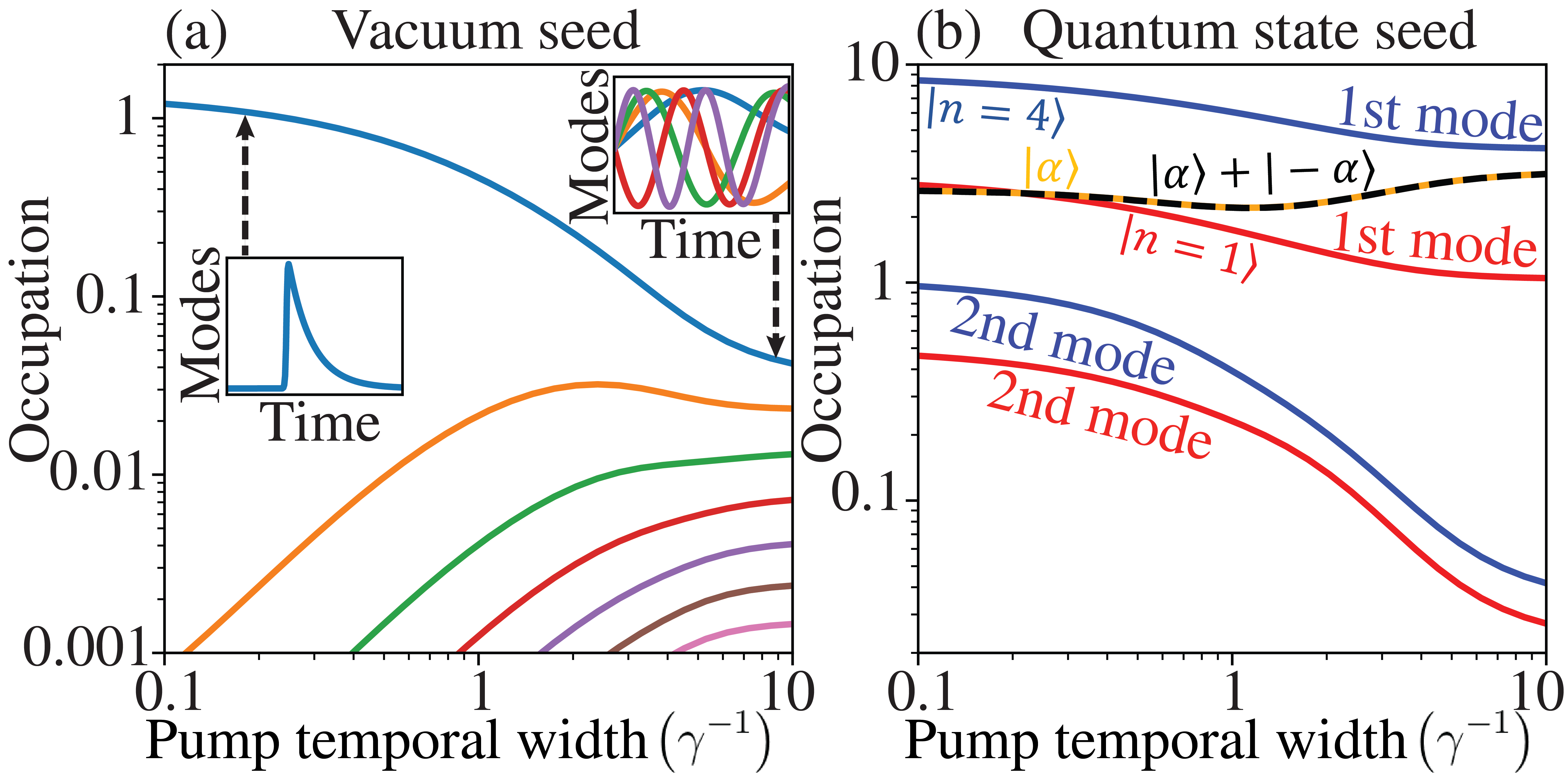}
\caption{\textbf{Temporal modes emitted by a parametric amplifier.} \textbf{(a)}  Vacuum input. Occupation of most populated modes as a function of pump duration for Gaussian pumps with constant pulse area  $(\int \xi(t) dt,\;\xi_0=1)$. A short pump pulse excites the cavity mode into a squeezed state that leaks into a single-mode travelling pulse. A longer pump pulse overlaps the emission process, and more modes become occupied. The mode shapes for the limiting cases of very wide and narrow pumps are plotted as insets. \textbf{(b)} Occupation of the modes fed by a Gaussian pump $(\xi_0=1.5)$ and different initial quantum states in the input pulse. The occupation of the modes seeded by an input pulse of duration $\tau=1/\gamma$ (two modes for Fock states, one mode for coherent and Schr\"odinger cat states) are plotted as a function of the pump width, assuming a constant pump pulse area. We assume that the pump and quantum pulses peak at the same time.}
    \label{Figure 2}
\end{figure}

We analyse first the output of the OPO with only vacuum seed and a time-dependent Gaussian pump $\xi(t)=\frac{\xi_0}{\sqrt{2\pi\sigma_\xi^2}}\exp({-\frac{(t-t_0)^2}{2\sigma_\xi^2}})$. Figure \ref{Figure 2}(a) shows the occupation of the modes of the output field. For a short pulsed pump, ($\gamma \sigma_\xi \ll 1$) only a single mode of the output field is excited (squeezed), as can be seen by the large occupation of the most occupied mode in the left-hand side of Figure \ref{Figure 2}(a). The mode occupied represents the exponential decay (shown in the inset) of the abruptly excited cavity mode. For wider pumps, the output field populates many modes (shown in the right-hand inset of Figure \ref{Figure 2}(a)). In the limiting case of an infinite duration constant pump, the photons possess strong frequency correlations, $\omega_1+\omega_2=\omega_{\textup{pump}}$, to conserve energy. In this limit, the output field explores a continuum of modes and $g^{(1)}(\omega_1,\omega_2)\propto S(\omega_1)\delta(\omega_\textup{pump}-\omega_1-\omega_2)$ where $S(\omega_1)$ is the output spectrum \cite{PhysRevA.30.1386}.

Next, we consider the parametric amplification of a quantum input pulse in the mode $u(t) \propto \exp\left(-\frac{(t-t_0)^2}{2\tau^2}\right)$ in Figure \ref{Figure 2}(b). Notice that the coherent state and the Schrödinger cat input feed only into one output mode as they obey the condition in Eq. \eqref{eq:single_mode_condition}. In contrast, Fock states lead to a population of two modes, as shown by the red and blue pairs of curves in Figure \ref{Figure 2}(b). The efficiency of the gain is largest for the case of a short pump. In this case, the emission by the cavity is negligible during the gain process, and the accumulated cavity field stimulates the strongest gain.


\begin{figure}
    \centering
    \includegraphics[width=\columnwidth]{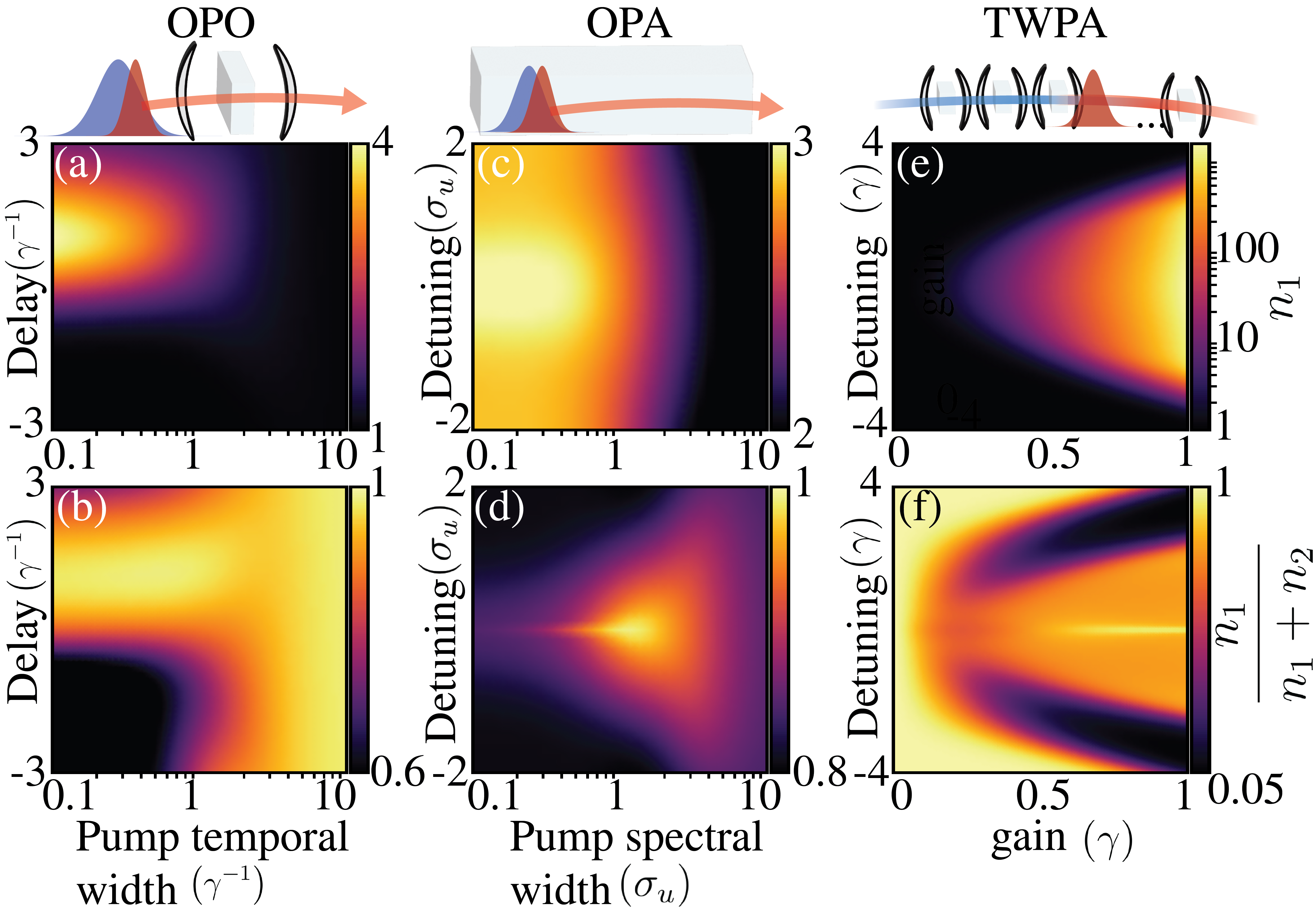}

\caption{\textbf{Comparison of three parametric amplifiers acting on a single photon input state.} Upper panels: Mean photon number in the dominant output mode $n_1$. Lower panels: occupation ratio $n_1/(n_1+n_2)$ as defined in Eq \eqref{eq:g1_fin1}. \textbf{(a-b)} For the OPO, these quantities are plotted as a function of the delay time between the pump and probe and the pump temporal width in units of the cavity decay rate $\gamma$. \textbf{(c-d)} For the OPA, with a spatially extended non-linear medium and no cavity, they are shown as a function of the pump-seed detuning and the pump spectral width in units of the spectral width of the input quantum pulse $\sigma_u$. \textbf{(e-f)} For the TWPA, they are shown as a function of the pump detuning in units of the cavity decay rate $\gamma$ and the total amount of gain for 3000 OPOs.}\label{Figure 3}
\end{figure}
\textit{Comparison of three parametric amplifiers--- }We recall that the output of the OPO contains both the transformed input quantum pulse(s) and components of the multi-mode squeezed vacuum. This results from our general analysis and applies to any setup with quadratic Hamiltonians. We shall consider exemplary systems and analyse the output field modes that are seeded by a single photon input pulse $u(t)\propto \exp\left(-\frac{t^2}{2\sigma_u^2}\right)$. The three setups are illustrated above the upper panels in Figure \ref{Figure 3}: 1. the Optical Parametric Oscillator (OPO) as explained above in Eq. \eqref{eq:OPO-Hamiltonian}, 2. The Optical Parametric Amplifier (OPA), with a spatially extended non-linear medium and no cavity, and 3. the Traveling Wave Parametric Amplifier (TWPA), represented here as a concatenated sequence of OPOs. For all three systems, we find regimes of close to ideal single-mode amplification, as shown by the regimes of large occupation in the dominant mode $v_1$ (upper panels in Figure \ref{Figure 3}) and the large ratio between the occupation in the dominant mode $n_1$ and the sum of the two most populated modes $n_1 + n_2$ (lower panels Figure \ref{Figure 3}). 

For the amplification of a single photon Gaussian pulse by the OPO, a short delay between the short pump and quantum state seed is optimal, for the OPA, a large overlap in frequency domain between the pump and seed will provide high gain and single mode operation, while for the TWPA, we find that resonant cavities provide the highest gain and highest single mode operation ratio. More information on the calculations presented in Figure \ref{Figure 3} is given in the SMV-VII. 

\begin{figure}
    \centering
\includegraphics[width=\columnwidth]{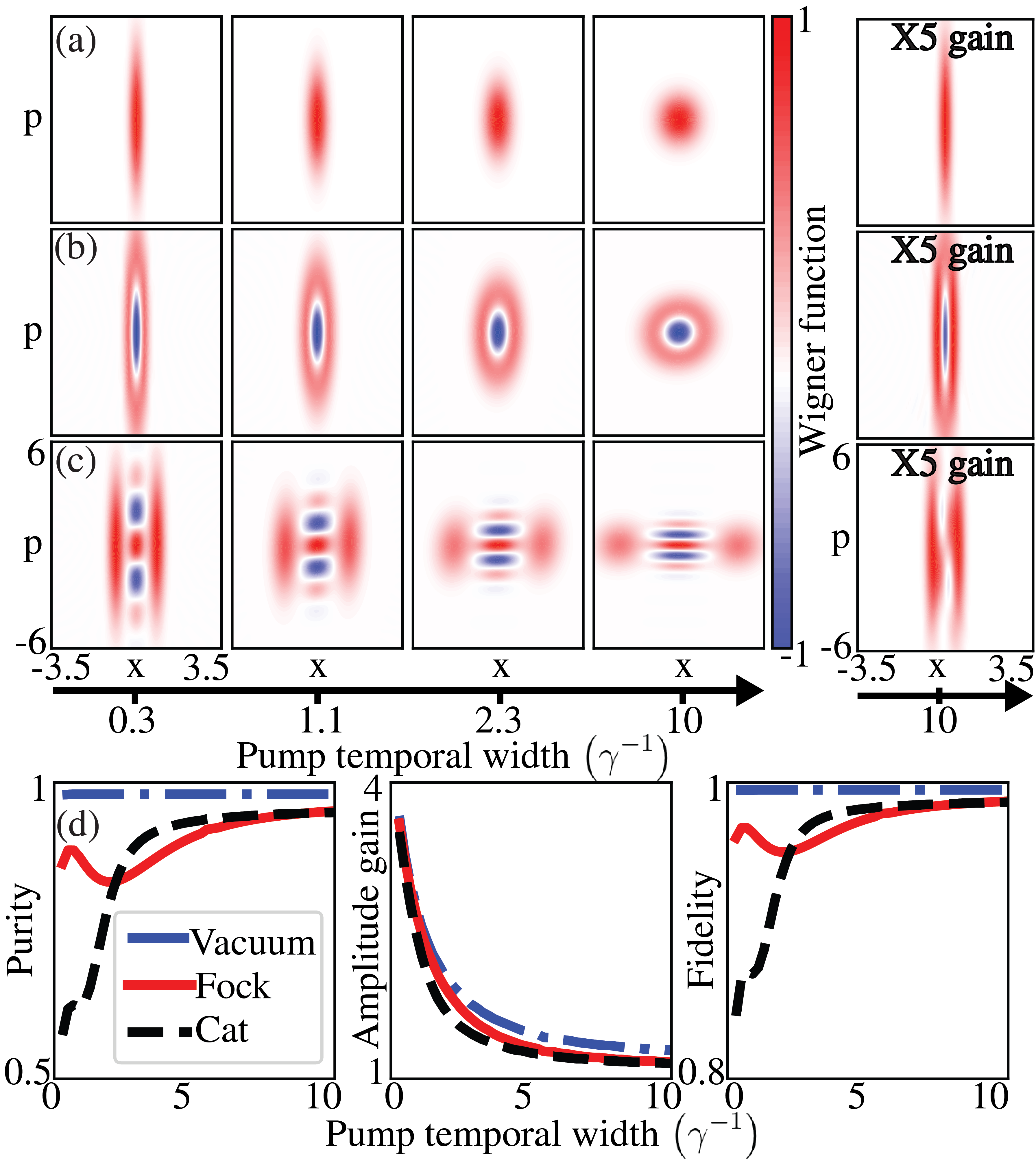}
\caption{\textbf{Transformation of quantum states by an OPO.} Wigner functions  are shown for the most occupied output temporal modes for different input states, \textbf{(a)} the vacuum state $\ket{0}$, \textbf{(b)} the single photon Fock state $\ket{n=1}$ and \textbf{(c)} the Schrödinger cat state $\ket{\alpha}+\ket{-\alpha}$ with $\alpha=2.5$. Assuming the same pump pulse area, short pulses lead to the strongest squeezing. \textbf{(d)} The purity of the state in the mode $v_1$ is plotted on the left. For each pump temporal width, we calculate the quantum state in the mode $v_1$, then we find the single mode parametrically amplified state to which it has the highest fidelity, we plot this fidelity and the corresponding squeezing on (d) (middle,right). The squeezing is quantified by the amplitude gain in the $p$ quadrature.}                 
\label{Figure 4}
\end{figure}

Next, we implement the theory to calculate the quantum state content of the most occupied mode. Results of these calculations are shown in Figure \ref{Figure 4} for squeezing by an OPO. The output Wigner functions in the most occupied single mode are plotted for different input states: vacuum (a), a single photon state (b), and a cat state (c) (squeezed cat states can be mixed to create GKP states \cite{GKP,PhysRevA.97.022341}). Results are shown for different pump widths, assuming the optimal delay identified in Figure \ref{Figure 3}. We observe the effect of parametric amplification for short pulses while, in agreement with Figure \ref{Figure 2}., for longer pump pulses, the quantum states experience little or no amplification. When the pump area is increased by a factor of 5 (rightmost column of Figure \ref{Figure 4}), the gain is stronger also for the longer pulses. However, in this regime the quantum states are polluted by the multi-mode squeezed vacuum generated by the OPO, decohering the quantum states.  

To make a quantitative comparison of the transformation of quantum pulses with the ideal single-mode squeezing transformation, we calculate the quantum state occupying the mode $v_1$, and in the leftmost panel, we plot the purity of this state, which can be lower than unity, both due to entanglement with the second output mode seeded by the input state $v_2$ and due to contamination by the squeezed vacuum modes $w_i$. In the rightmost panel, we compare the state in $v_1$ to the input state squeezed by the single mode transformation of Eq. \eqref{single_mode_Bogolyubov}, optimizing over the amount of squeezing, parameterized as gain in the p quadrature, which we plot in the central panel of Figure \ref{Figure 4}(d).
Notably, we find pure state fidelities of over 85\% with more than a factor 3 of parametric amplification on the $p$-quadrature, while close to unity fidelity parametric amplification of a single mode pulse is only possible for much smaller gains, except for the special case of vacuum input. We note that we considered an input seed and pump in a Gaussian temporal mode and that optimizing over these may increase the fidelity, purity and squeezing. 

\textit{Conclusion---}
An optical component described by a multi-mode field Hamiltonian that is quadratic in creation and annihilation operators causes linear dispersion of wave packets combined with correlated creation and annihilation of photon pairs over all modes. A multi-mode Bogoliubov transformation accounts for the effect of the Hamiltonian on the input field operators. We have shown that if all but a single mode of the input field are in the vacuum state, despite the multi-mode character of the problem, only two output modes will contain quantum states that depend on the non-trivial input state. These output modes will generally also contain components of squeezed vacuum from the parametric amplifier. While these results are at variance with the ideal parametric amplification as a unitary operation on a single-mode quantum field, our analysis can identify optimal parameter settings for reducing additional noise and spreading over more than a single mode. 

Our theory applies also to multiple input and output modes, and the joint quantum state of two or more modes can be readily found. Optimization of the fidelity of the parametric amplification process and the more general properties of the output quantum state of the multi-mode Bogoliubov transformation constitute a promising topic for further study.

V. Rueskov Christiansen acknowledges support from the Danish National Research Foundation through the Center of Excellence for Complex Quantum Systems (Grant agreement No. DNRF156) and discussions with Daniel Holleufer.
K. M\o lmer acknowledges support from the Carlsberg Foundation through the ``Semper Ardens'' Research Project QCooL.

\bibliography{bibliography.bib}

\begin{thebibliography}{30}%
\makeatletter
\providecommand \@ifxundefined [1]{%
 \@ifx{#1\undefined}
}%
\providecommand \@ifnum [1]{%
 \ifnum #1\expandafter \@firstoftwo
 \else \expandafter \@secondoftwo
 \fi
}%
\providecommand \@ifx [1]{%
 \ifx #1\expandafter \@firstoftwo
 \else \expandafter \@secondoftwo
 \fi
}%
\providecommand \natexlab [1]{#1}%
\providecommand \enquote  [1]{``#1''}%
\providecommand \bibnamefont  [1]{#1}%
\providecommand \bibfnamefont [1]{#1}%
\providecommand \citenamefont [1]{#1}%
\providecommand \href@noop [0]{\@secondoftwo}%
\providecommand \href [0]{\begingroup \@sanitize@url \@href}%
\providecommand \@href[1]{\@@startlink{#1}\@@href}%
\providecommand \@@href[1]{\endgroup#1\@@endlink}%
\providecommand \@sanitize@url [0]{\catcode `\\12\catcode `\$12\catcode `\&12\catcode `\#12\catcode `\^12\catcode `\_12\catcode `\%12\relax}%
\providecommand \@@startlink[1]{}%
\providecommand \@@endlink[0]{}%
\providecommand \url  [0]{\begingroup\@sanitize@url \@url }%
\providecommand \@url [1]{\endgroup\@href {#1}{\urlprefix }}%
\providecommand \urlprefix  [0]{URL }%
\providecommand \Eprint [0]{\href }%
\providecommand \doibase [0]{https://doi.org/}%
\providecommand \selectlanguage [0]{\@gobble}%
\providecommand \bibinfo  [0]{\@secondoftwo}%
\providecommand \bibfield  [0]{\@secondoftwo}%
\providecommand \translation [1]{[#1]}%
\providecommand \BibitemOpen [0]{}%
\providecommand \bibitemStop [0]{}%
\providecommand \bibitemNoStop [0]{.\EOS\space}%
\providecommand \EOS [0]{\spacefactor3000\relax}%
\providecommand \BibitemShut  [1]{\csname bibitem#1\endcsname}%
\let\auto@bib@innerbib\@empty
\bibitem [{\citenamefont {O'Brien}\ \emph {et~al.}(2009)\citenamefont {O'Brien}, \citenamefont {Furusawa},\ and\ \citenamefont {Vu{\v c}kovi{\'c}}}]{photonic_quantum_tech}%
  \BibitemOpen
  \bibfield  {author} {\bibinfo {author} {\bibfnamefont {J.~L.}\ \bibnamefont {O'Brien}}, \bibinfo {author} {\bibfnamefont {A.}~\bibnamefont {Furusawa}},\ and\ \bibinfo {author} {\bibfnamefont {J.}~\bibnamefont {Vu{\v c}kovi{\'c}}},\ }\bibfield  {title} {\bibinfo {title} {Photonic quantum technologies},\ }\href {https://doi.org/10.1038/nphoton.2009.229} {\bibfield  {journal} {\bibinfo  {journal} {Nature Photonics}\ }\textbf {\bibinfo {volume} {3}},\ \bibinfo {pages} {687} (\bibinfo {year} {2009})}\BibitemShut {NoStop}%
\bibitem [{\citenamefont {Kimble}(2008)}]{internet_quantum}%
  \BibitemOpen
  \bibfield  {author} {\bibinfo {author} {\bibfnamefont {H.~J.}\ \bibnamefont {Kimble}},\ }\bibfield  {title} {\bibinfo {title} {The quantum internet},\ }\href {https://doi.org/10.1038/nature07127} {\bibfield  {journal} {\bibinfo  {journal} {Nature}\ }\textbf {\bibinfo {volume} {453}},\ \bibinfo {pages} {1023} (\bibinfo {year} {2008})}\BibitemShut {NoStop}%
\bibitem [{\citenamefont {Korzh}\ \emph {et~al.}(2015)\citenamefont {Korzh}, \citenamefont {Lim}, \citenamefont {Houlmann}, \citenamefont {Gisin}, \citenamefont {Li}, \citenamefont {Nolan}, \citenamefont {Sanguinetti}, \citenamefont {Thew},\ and\ \citenamefont {Zbinden}}]{QKD}%
  \BibitemOpen
  \bibfield  {author} {\bibinfo {author} {\bibfnamefont {B.}~\bibnamefont {Korzh}}, \bibinfo {author} {\bibfnamefont {C.~C.~W.}\ \bibnamefont {Lim}}, \bibinfo {author} {\bibfnamefont {R.}~\bibnamefont {Houlmann}}, \bibinfo {author} {\bibfnamefont {N.}~\bibnamefont {Gisin}}, \bibinfo {author} {\bibfnamefont {M.~J.}\ \bibnamefont {Li}}, \bibinfo {author} {\bibfnamefont {D.}~\bibnamefont {Nolan}}, \bibinfo {author} {\bibfnamefont {B.}~\bibnamefont {Sanguinetti}}, \bibinfo {author} {\bibfnamefont {R.}~\bibnamefont {Thew}},\ and\ \bibinfo {author} {\bibfnamefont {H.}~\bibnamefont {Zbinden}},\ }\bibfield  {title} {\bibinfo {title} {Provably secure and practical quantum key distribution over 307 km of optical fibre},\ }\href {https://doi.org/10.1038/nphoton.2014.327} {\bibfield  {journal} {\bibinfo  {journal} {Nature Photonics}\ }\textbf {\bibinfo {volume} {9}},\ \bibinfo {pages} {163} (\bibinfo {year} {2015})}\BibitemShut {NoStop}%
\bibitem [{\citenamefont {Caves}(1982)}]{PhysRevD.26.1817}%
  \BibitemOpen
  \bibfield  {author} {\bibinfo {author} {\bibfnamefont {C.~M.}\ \bibnamefont {Caves}},\ }\bibfield  {title} {\bibinfo {title} {Quantum limits on noise in linear amplifiers},\ }\href {https://doi.org/10.1103/PhysRevD.26.1817} {\bibfield  {journal} {\bibinfo  {journal} {Phys. Rev. D}\ }\textbf {\bibinfo {volume} {26}},\ \bibinfo {pages} {1817} (\bibinfo {year} {1982})}\BibitemShut {NoStop}%
\bibitem [{\citenamefont {Caves}(2020)}]{cavesnew}%
  \BibitemOpen
  \bibfield  {author} {\bibinfo {author} {\bibfnamefont {C.~M.}\ \bibnamefont {Caves}},\ }\bibfield  {title} {\bibinfo {title} {Reframing su(1,1) interferometry},\ }\href {https://doi.org/https://doi.org/10.1002/qute.201900138} {\bibfield  {journal} {\bibinfo  {journal} {Advanced Quantum Technologies}\ }\textbf {\bibinfo {volume} {3}},\ \bibinfo {pages} {1900138} (\bibinfo {year} {2020})}\BibitemShut {NoStop}%
\bibitem [{\citenamefont {Lawrie}\ \emph {et~al.}(2019)\citenamefont {Lawrie}, \citenamefont {Lett}, \citenamefont {Marino},\ and\ \citenamefont {Pooser}}]{sensing}%
  \BibitemOpen
  \bibfield  {author} {\bibinfo {author} {\bibfnamefont {B.~J.}\ \bibnamefont {Lawrie}}, \bibinfo {author} {\bibfnamefont {P.~D.}\ \bibnamefont {Lett}}, \bibinfo {author} {\bibfnamefont {A.~M.}\ \bibnamefont {Marino}},\ and\ \bibinfo {author} {\bibfnamefont {R.~C.}\ \bibnamefont {Pooser}},\ }\bibfield  {title} {\bibinfo {title} {Quantum sensing with squeezed light},\ }\bibfield  {journal} {\bibinfo  {journal} {ACS Photonics}\ }\href {https://doi.org/10.1021/acsphotonics.9b00250} {10.1021/acsphotonics.9b00250} (\bibinfo {year} {2019})\BibitemShut {NoStop}%
\bibitem [{\citenamefont {Gottesman}\ \emph {et~al.}(2001)\citenamefont {Gottesman}, \citenamefont {Kitaev},\ and\ \citenamefont {Preskill}}]{GKP}%
  \BibitemOpen
  \bibfield  {author} {\bibinfo {author} {\bibfnamefont {D.}~\bibnamefont {Gottesman}}, \bibinfo {author} {\bibfnamefont {A.}~\bibnamefont {Kitaev}},\ and\ \bibinfo {author} {\bibfnamefont {J.}~\bibnamefont {Preskill}},\ }\bibfield  {title} {\bibinfo {title} {Encoding a qubit in an oscillator},\ }\href {https://doi.org/10.1103/PhysRevA.64.012310} {\bibfield  {journal} {\bibinfo  {journal} {Phys. Rev. A}\ }\textbf {\bibinfo {volume} {64}},\ \bibinfo {pages} {012310} (\bibinfo {year} {2001})}\BibitemShut {NoStop}%
\bibitem [{\citenamefont {Sivak}\ \emph {et~al.}(2023)\citenamefont {Sivak}, \citenamefont {Eickbusch}, \citenamefont {Royer}, \citenamefont {Singh}, \citenamefont {Tsioutsios}, \citenamefont {Ganjam}, \citenamefont {Miano}, \citenamefont {Brock}, \citenamefont {Ding}, \citenamefont {Frunzio}, \citenamefont {Girvin}, \citenamefont {Schoelkopf},\ and\ \citenamefont {Devoret}}]{GKP_exp}%
  \BibitemOpen
  \bibfield  {author} {\bibinfo {author} {\bibfnamefont {V.~V.}\ \bibnamefont {Sivak}}, \bibinfo {author} {\bibfnamefont {A.}~\bibnamefont {Eickbusch}}, \bibinfo {author} {\bibfnamefont {B.}~\bibnamefont {Royer}}, \bibinfo {author} {\bibfnamefont {S.}~\bibnamefont {Singh}}, \bibinfo {author} {\bibfnamefont {I.}~\bibnamefont {Tsioutsios}}, \bibinfo {author} {\bibfnamefont {S.}~\bibnamefont {Ganjam}}, \bibinfo {author} {\bibfnamefont {A.}~\bibnamefont {Miano}}, \bibinfo {author} {\bibfnamefont {B.~L.}\ \bibnamefont {Brock}}, \bibinfo {author} {\bibfnamefont {A.~Z.}\ \bibnamefont {Ding}}, \bibinfo {author} {\bibfnamefont {L.}~\bibnamefont {Frunzio}}, \bibinfo {author} {\bibfnamefont {S.~M.}\ \bibnamefont {Girvin}}, \bibinfo {author} {\bibfnamefont {R.~J.}\ \bibnamefont {Schoelkopf}},\ and\ \bibinfo {author} {\bibfnamefont {M.~H.}\ \bibnamefont {Devoret}},\ }\bibfield  {title} {\bibinfo {title} {Real-time quantum error correction beyond break-even},\ }\href {https://doi.org/10.1038/s41586-023-05782-6} {\bibfield
  {journal} {\bibinfo  {journal} {Nature}\ }\textbf {\bibinfo {volume} {616}},\ \bibinfo {pages} {50} (\bibinfo {year} {2023})}\BibitemShut {NoStop}%
\bibitem [{\citenamefont {Briegel}\ \emph {et~al.}(2009)\citenamefont {Briegel}, \citenamefont {Browne}, \citenamefont {D{\"u}r}, \citenamefont {Raussendorf},\ and\ \citenamefont {Van~den Nest}}]{measurment_based}%
  \BibitemOpen
  \bibfield  {author} {\bibinfo {author} {\bibfnamefont {H.~J.}\ \bibnamefont {Briegel}}, \bibinfo {author} {\bibfnamefont {D.~E.}\ \bibnamefont {Browne}}, \bibinfo {author} {\bibfnamefont {W.}~\bibnamefont {D{\"u}r}}, \bibinfo {author} {\bibfnamefont {R.}~\bibnamefont {Raussendorf}},\ and\ \bibinfo {author} {\bibfnamefont {M.}~\bibnamefont {Van~den Nest}},\ }\bibfield  {title} {\bibinfo {title} {Measurement-based quantum computation},\ }\href {https://doi.org/10.1038/nphys1157} {\bibfield  {journal} {\bibinfo  {journal} {Nature Physics}\ }\textbf {\bibinfo {volume} {5}},\ \bibinfo {pages} {19} (\bibinfo {year} {2009})}\BibitemShut {NoStop}%
\bibitem [{\citenamefont {Fabre}\ and\ \citenamefont {Treps}(2020)}]{RevModPhys.92.035005}%
  \BibitemOpen
  \bibfield  {author} {\bibinfo {author} {\bibfnamefont {C.}~\bibnamefont {Fabre}}\ and\ \bibinfo {author} {\bibfnamefont {N.}~\bibnamefont {Treps}},\ }\bibfield  {title} {\bibinfo {title} {Modes and states in quantum optics},\ }\href {https://doi.org/10.1103/RevModPhys.92.035005} {\bibfield  {journal} {\bibinfo  {journal} {Rev. Mod. Phys.}\ }\textbf {\bibinfo {volume} {92}},\ \bibinfo {pages} {035005} (\bibinfo {year} {2020})}\BibitemShut {NoStop}%
\bibitem [{\citenamefont {Caves}(1981)}]{PhysRevD.23.1693}%
  \BibitemOpen
  \bibfield  {author} {\bibinfo {author} {\bibfnamefont {C.~M.}\ \bibnamefont {Caves}},\ }\bibfield  {title} {\bibinfo {title} {Quantum-mechanical noise in an interferometer},\ }\href {https://doi.org/10.1103/PhysRevD.23.1693} {\bibfield  {journal} {\bibinfo  {journal} {Phys. Rev. D}\ }\textbf {\bibinfo {volume} {23}},\ \bibinfo {pages} {1693} (\bibinfo {year} {1981})}\BibitemShut {NoStop}%
\bibitem [{\citenamefont {Kalash}\ and\ \citenamefont {Chekhova}(2023)}]{Kalash:23}%
  \BibitemOpen
  \bibfield  {author} {\bibinfo {author} {\bibfnamefont {M.}~\bibnamefont {Kalash}}\ and\ \bibinfo {author} {\bibfnamefont {M.~V.}\ \bibnamefont {Chekhova}},\ }\bibfield  {title} {\bibinfo {title} {Wigner function tomography via optical parametric amplification},\ }\href {https://doi.org/10.1364/OPTICA.488697} {\bibfield  {journal} {\bibinfo  {journal} {Optica}\ }\textbf {\bibinfo {volume} {10}},\ \bibinfo {pages} {1142} (\bibinfo {year} {2023})}\BibitemShut {NoStop}%
\bibitem [{\citenamefont {Li}\ \emph {et~al.}(2017)\citenamefont {Li}, \citenamefont {Anderson}, \citenamefont {Horrom}, \citenamefont {Schmittberger}, \citenamefont {Jones},\ and\ \citenamefont {Lett}}]{Li:17}%
  \BibitemOpen
  \bibfield  {author} {\bibinfo {author} {\bibfnamefont {T.}~\bibnamefont {Li}}, \bibinfo {author} {\bibfnamefont {B.~E.}\ \bibnamefont {Anderson}}, \bibinfo {author} {\bibfnamefont {T.}~\bibnamefont {Horrom}}, \bibinfo {author} {\bibfnamefont {B.~L.}\ \bibnamefont {Schmittberger}}, \bibinfo {author} {\bibfnamefont {K.~M.}\ \bibnamefont {Jones}},\ and\ \bibinfo {author} {\bibfnamefont {P.~D.}\ \bibnamefont {Lett}},\ }\bibfield  {title} {\bibinfo {title} {Improved measurement of two-mode quantum correlations using a phase-sensitive amplifier},\ }\href {https://doi.org/10.1364/OE.25.021301} {\bibfield  {journal} {\bibinfo  {journal} {Opt. Express}\ }\textbf {\bibinfo {volume} {25}},\ \bibinfo {pages} {21301} (\bibinfo {year} {2017})}\BibitemShut {NoStop}%
\bibitem [{\citenamefont {Macklin}\ \emph {et~al.}(2015)\citenamefont {Macklin}, \citenamefont {O'Brien}, \citenamefont {Hover}, \citenamefont {Schwartz}, \citenamefont {Bolkhovsky}, \citenamefont {Zhang}, \citenamefont {Oliver},\ and\ \citenamefont {Siddiqi}}]{TWPA_science_2015}%
  \BibitemOpen
  \bibfield  {author} {\bibinfo {author} {\bibfnamefont {C.}~\bibnamefont {Macklin}}, \bibinfo {author} {\bibfnamefont {K.}~\bibnamefont {O'Brien}}, \bibinfo {author} {\bibfnamefont {D.}~\bibnamefont {Hover}}, \bibinfo {author} {\bibfnamefont {M.~E.}\ \bibnamefont {Schwartz}}, \bibinfo {author} {\bibfnamefont {V.}~\bibnamefont {Bolkhovsky}}, \bibinfo {author} {\bibfnamefont {X.}~\bibnamefont {Zhang}}, \bibinfo {author} {\bibfnamefont {W.~D.}\ \bibnamefont {Oliver}},\ and\ \bibinfo {author} {\bibfnamefont {I.}~\bibnamefont {Siddiqi}},\ }\bibfield  {title} {\bibinfo {title} {A near-quantum-limited josephson traveling-wave parametric amplifier},\ }\href {https://doi.org/10.1126/science.aaa8525} {\bibfield  {journal} {\bibinfo  {journal} {Science}\ }\textbf {\bibinfo {volume} {350}},\ \bibinfo {pages} {307} (\bibinfo {year} {2015})}\BibitemShut {NoStop}%
\bibitem [{\citenamefont {Qiu}\ \emph {et~al.}(2023)\citenamefont {Qiu}, \citenamefont {Grimsmo}, \citenamefont {Peng}, \citenamefont {Kannan}, \citenamefont {Lienhard}, \citenamefont {Sung}, \citenamefont {Krantz}, \citenamefont {Bolkhovsky}, \citenamefont {Calusine}, \citenamefont {Kim}, \citenamefont {Melville}, \citenamefont {Niedzielski}, \citenamefont {Yoder}, \citenamefont {Schwartz}, \citenamefont {Orlando}, \citenamefont {Siddiqi}, \citenamefont {Gustavsson}, \citenamefont {O'Brien},\ and\ \citenamefont {Oliver}}]{TWPA_2023}%
  \BibitemOpen
  \bibfield  {author} {\bibinfo {author} {\bibfnamefont {J.~Y.}\ \bibnamefont {Qiu}}, \bibinfo {author} {\bibfnamefont {A.}~\bibnamefont {Grimsmo}}, \bibinfo {author} {\bibfnamefont {K.}~\bibnamefont {Peng}}, \bibinfo {author} {\bibfnamefont {B.}~\bibnamefont {Kannan}}, \bibinfo {author} {\bibfnamefont {B.}~\bibnamefont {Lienhard}}, \bibinfo {author} {\bibfnamefont {Y.}~\bibnamefont {Sung}}, \bibinfo {author} {\bibfnamefont {P.}~\bibnamefont {Krantz}}, \bibinfo {author} {\bibfnamefont {V.}~\bibnamefont {Bolkhovsky}}, \bibinfo {author} {\bibfnamefont {G.}~\bibnamefont {Calusine}}, \bibinfo {author} {\bibfnamefont {D.}~\bibnamefont {Kim}}, \bibinfo {author} {\bibfnamefont {A.}~\bibnamefont {Melville}}, \bibinfo {author} {\bibfnamefont {B.~M.}\ \bibnamefont {Niedzielski}}, \bibinfo {author} {\bibfnamefont {J.}~\bibnamefont {Yoder}}, \bibinfo {author} {\bibfnamefont {M.~E.}\ \bibnamefont {Schwartz}}, \bibinfo {author} {\bibfnamefont {T.~P.}\ \bibnamefont {Orlando}}, \bibinfo {author} {\bibfnamefont
  {I.}~\bibnamefont {Siddiqi}}, \bibinfo {author} {\bibfnamefont {S.}~\bibnamefont {Gustavsson}}, \bibinfo {author} {\bibfnamefont {K.~P.}\ \bibnamefont {O'Brien}},\ and\ \bibinfo {author} {\bibfnamefont {W.~D.}\ \bibnamefont {Oliver}},\ }\bibfield  {title} {\bibinfo {title} {Broadband squeezed microwaves and amplification with a josephson travelling-wave parametric amplifier},\ }\href {https://doi.org/10.1038/s41567-022-01929-w} {\bibfield  {journal} {\bibinfo  {journal} {Nature Physics}\ }\textbf {\bibinfo {volume} {19}},\ \bibinfo {pages} {706} (\bibinfo {year} {2023})}\BibitemShut {NoStop}%
\bibitem [{\citenamefont {Nehra}\ \emph {et~al.}(2022)\citenamefont {Nehra}, \citenamefont {Sekine}, \citenamefont {Ledezma}, \citenamefont {Guo}, \citenamefont {Gray}, \citenamefont {Roy},\ and\ \citenamefont {Marandi}}]{doi:10.1126/science.abo6213}%
  \BibitemOpen
  \bibfield  {author} {\bibinfo {author} {\bibfnamefont {R.}~\bibnamefont {Nehra}}, \bibinfo {author} {\bibfnamefont {R.}~\bibnamefont {Sekine}}, \bibinfo {author} {\bibfnamefont {L.}~\bibnamefont {Ledezma}}, \bibinfo {author} {\bibfnamefont {Q.}~\bibnamefont {Guo}}, \bibinfo {author} {\bibfnamefont {R.~M.}\ \bibnamefont {Gray}}, \bibinfo {author} {\bibfnamefont {A.}~\bibnamefont {Roy}},\ and\ \bibinfo {author} {\bibfnamefont {A.}~\bibnamefont {Marandi}},\ }\bibfield  {title} {\bibinfo {title} {Few-cycle vacuum squeezing in nanophotonics},\ }\href {https://doi.org/10.1126/science.abo6213} {\bibfield  {journal} {\bibinfo  {journal} {Science}\ }\textbf {\bibinfo {volume} {377}},\ \bibinfo {pages} {1333} (\bibinfo {year} {2022})}\BibitemShut {NoStop}%
\bibitem [{\citenamefont {Lloyd}\ and\ \citenamefont {Braunstein}(1999)}]{PhysRevLett.82.1784}%
  \BibitemOpen
  \bibfield  {author} {\bibinfo {author} {\bibfnamefont {S.}~\bibnamefont {Lloyd}}\ and\ \bibinfo {author} {\bibfnamefont {S.~L.}\ \bibnamefont {Braunstein}},\ }\bibfield  {title} {\bibinfo {title} {Quantum computation over continuous variables},\ }\href {https://doi.org/10.1103/PhysRevLett.82.1784} {\bibfield  {journal} {\bibinfo  {journal} {Phys. Rev. Lett.}\ }\textbf {\bibinfo {volume} {82}},\ \bibinfo {pages} {1784} (\bibinfo {year} {1999})}\BibitemShut {NoStop}%
\bibitem [{\citenamefont {Weigand}\ and\ \citenamefont {Terhal}(2018)}]{PhysRevA.97.022341}%
  \BibitemOpen
  \bibfield  {author} {\bibinfo {author} {\bibfnamefont {D.~J.}\ \bibnamefont {Weigand}}\ and\ \bibinfo {author} {\bibfnamefont {B.~M.}\ \bibnamefont {Terhal}},\ }\bibfield  {title} {\bibinfo {title} {Generating grid states from schr\"odinger-cat states without postselection},\ }\href {https://doi.org/10.1103/PhysRevA.97.022341} {\bibfield  {journal} {\bibinfo  {journal} {Phys. Rev. A}\ }\textbf {\bibinfo {volume} {97}},\ \bibinfo {pages} {022341} (\bibinfo {year} {2018})}\BibitemShut {NoStop}%
\bibitem [{\citenamefont {Yanagimoto}\ \emph {et~al.}(2023)\citenamefont {Yanagimoto}, \citenamefont {Nehra}, \citenamefont {Hamerly}, \citenamefont {Ng}, \citenamefont {Marandi},\ and\ \citenamefont {Mabuchi}}]{GKP_OPA_singlemode_approx}%
  \BibitemOpen
  \bibfield  {author} {\bibinfo {author} {\bibfnamefont {R.}~\bibnamefont {Yanagimoto}}, \bibinfo {author} {\bibfnamefont {R.}~\bibnamefont {Nehra}}, \bibinfo {author} {\bibfnamefont {R.}~\bibnamefont {Hamerly}}, \bibinfo {author} {\bibfnamefont {E.}~\bibnamefont {Ng}}, \bibinfo {author} {\bibfnamefont {A.}~\bibnamefont {Marandi}},\ and\ \bibinfo {author} {\bibfnamefont {H.}~\bibnamefont {Mabuchi}},\ }\bibfield  {title} {\bibinfo {title} {Quantum nondemolition measurements with optical parametric amplifiers for ultrafast universal quantum information processing},\ }\href {https://doi.org/10.1103/PRXQuantum.4.010333} {\bibfield  {journal} {\bibinfo  {journal} {PRX Quantum}\ }\textbf {\bibinfo {volume} {4}},\ \bibinfo {pages} {010333} (\bibinfo {year} {2023})}\BibitemShut {NoStop}%
\bibitem [{sup()}]{supp}%
  \BibitemOpen
  \href@noop {} {\bibinfo  {journal} {See supplemental material at (url to be inserted by publisher) for derivations of central results}\ }\BibitemShut {NoStop}%
\bibitem [{\citenamefont {Christ}\ \emph {et~al.}(2013)\citenamefont {Christ}, \citenamefont {Brecht}, \citenamefont {Mauerer},\ and\ \citenamefont {Silberhorn}}]{Christ_2013}%
  \BibitemOpen
\bibfield  {journal} {  }\bibfield  {author} {\bibinfo {author} {\bibfnamefont {A.}~\bibnamefont {Christ}}, \bibinfo {author} {\bibfnamefont {B.}~\bibnamefont {Brecht}}, \bibinfo {author} {\bibfnamefont {W.}~\bibnamefont {Mauerer}},\ and\ \bibinfo {author} {\bibfnamefont {C.}~\bibnamefont {Silberhorn}},\ }\bibfield  {title} {\bibinfo {title} {Theory of quantum frequency conversion and type-ii parametric down-conversion in the high-gain regime},\ }\href {https://doi.org/10.1088/1367-2630/15/5/053038} {\bibfield  {journal} {\bibinfo  {journal} {New Journal of Physics}\ }\textbf {\bibinfo {volume} {15}},\ \bibinfo {pages} {053038} (\bibinfo {year} {2013})}\BibitemShut {NoStop}%
\bibitem [{\citenamefont {Bogolyubov}(1958)}]{Bogolyubov:1958se}%
  \BibitemOpen
  \bibfield  {author} {\bibinfo {author} {\bibfnamefont {N.~N.}\ \bibnamefont {Bogolyubov}},\ }\bibfield  {title} {\bibinfo {title} {{A New method in the theory of superconductivity. I}},\ }\href@noop {} {\bibfield  {journal} {\bibinfo  {journal} {Sov. Phys. JETP}\ }\textbf {\bibinfo {volume} {7}},\ \bibinfo {pages} {41} (\bibinfo {year} {1958})}\BibitemShut {NoStop}%
\bibitem [{\citenamefont {Aaronson}\ and\ \citenamefont {Arkhipov}(2011)}]{10.1145/1993636.1993682}%
  \BibitemOpen
  \bibfield  {author} {\bibinfo {author} {\bibfnamefont {S.}~\bibnamefont {Aaronson}}\ and\ \bibinfo {author} {\bibfnamefont {A.}~\bibnamefont {Arkhipov}},\ }\bibfield  {title} {\bibinfo {title} {The computational complexity of linear optics},\ }in\ \href {https://doi.org/10.1145/1993636.1993682} {\emph {\bibinfo {booktitle} {Proceedings of the Forty-Third Annual ACM Symposium on Theory of Computing}}},\ \bibinfo {series and number} {STOC '11}\ (\bibinfo  {publisher} {Association for Computing Machinery},\ \bibinfo {address} {New York, NY, USA},\ \bibinfo {year} {2011})\ p.\ \bibinfo {pages} {333–342}\BibitemShut {NoStop}%
\bibitem [{\citenamefont {Braunstein}(2005)}]{PhysRevA.71.055801}%
  \BibitemOpen
  \bibfield  {author} {\bibinfo {author} {\bibfnamefont {S.~L.}\ \bibnamefont {Braunstein}},\ }\bibfield  {title} {\bibinfo {title} {Squeezing as an irreducible resource},\ }\href {https://doi.org/10.1103/PhysRevA.71.055801} {\bibfield  {journal} {\bibinfo  {journal} {Phys. Rev. A}\ }\textbf {\bibinfo {volume} {71}},\ \bibinfo {pages} {055801} (\bibinfo {year} {2005})}\BibitemShut {NoStop}%
\bibitem [{\citenamefont {Cariolaro}\ and\ \citenamefont {Pierobon}(2016)}]{spec_Bloch_messiah}%
  \BibitemOpen
  \bibfield  {author} {\bibinfo {author} {\bibfnamefont {G.}~\bibnamefont {Cariolaro}}\ and\ \bibinfo {author} {\bibfnamefont {G.}~\bibnamefont {Pierobon}},\ }\bibfield  {title} {\bibinfo {title} {Reexamination of bloch-messiah reduction},\ }\href {https://doi.org/10.1103/PhysRevA.93.062115} {\bibfield  {journal} {\bibinfo  {journal} {Phys. Rev. A}\ }\textbf {\bibinfo {volume} {93}},\ \bibinfo {pages} {062115} (\bibinfo {year} {2016})}\BibitemShut {NoStop}%
\bibitem [{\citenamefont {Ekert}\ and\ \citenamefont {Knight}(1989)}]{Ekert}%
  \BibitemOpen
  \bibfield  {author} {\bibinfo {author} {\bibfnamefont {A.~K.}\ \bibnamefont {Ekert}}\ and\ \bibinfo {author} {\bibfnamefont {P.~L.}\ \bibnamefont {Knight}},\ }\bibfield  {title} {\bibinfo {title} {The wigner function of two-mode squeezed states; free and dissipative evolution},\ }in\ \href@noop {} {\emph {\bibinfo {booktitle} {Coherence and Quantum Optics VI}}},\ \bibinfo {editor} {edited by\ \bibinfo {editor} {\bibfnamefont {J.~H.}\ \bibnamefont {Eberly}}, \bibinfo {editor} {\bibfnamefont {L.}~\bibnamefont {Mandel}},\ and\ \bibinfo {editor} {\bibfnamefont {E.}~\bibnamefont {Wolf}}}\ (\bibinfo  {publisher} {Springer US},\ \bibinfo {address} {Boston, MA},\ \bibinfo {year} {1989})\ pp.\ \bibinfo {pages} {255--259}\BibitemShut {NoStop}%
\bibitem [{Note1()}]{Note1}%
  \BibitemOpen
  \bibinfo {note} {Link to the repository with Python code to find the output quantum state for a given transformation https://github.com/offektziperman/Amplifying-a-quantum-pulse}\BibitemShut {NoStop}%
\bibitem [{Note2()}]{Note2}%
  \BibitemOpen
  \bibinfo {note} {This input-output relation assumes unidirectional propagation of the field which may be ensured by a ring cavity, a circulator, or the phase matching in the OPA gain material in the absence of a cavity.}\BibitemShut {Stop}%
\bibitem [{\citenamefont {Gardiner}\ and\ \citenamefont {Collett}(1985)}]{Input-Output-Theory}%
  \BibitemOpen
  \bibfield  {author} {\bibinfo {author} {\bibfnamefont {C.~W.}\ \bibnamefont {Gardiner}}\ and\ \bibinfo {author} {\bibfnamefont {M.~J.}\ \bibnamefont {Collett}},\ }\bibfield  {title} {\bibinfo {title} {Input and output in damped quantum systems: Quantum stochastic differential equations and the master equation},\ }\href {https://doi.org/10.1103/PhysRevA.31.3761} {\bibfield  {journal} {\bibinfo  {journal} {Phys. Rev. A}\ }\textbf {\bibinfo {volume} {31}},\ \bibinfo {pages} {3761} (\bibinfo {year} {1985})}\BibitemShut {NoStop}%
\bibitem [{\citenamefont {Collett}\ and\ \citenamefont {Gardiner}(1984)}]{PhysRevA.30.1386}%
  \BibitemOpen
  \bibfield  {author} {\bibinfo {author} {\bibfnamefont {M.~J.}\ \bibnamefont {Collett}}\ and\ \bibinfo {author} {\bibfnamefont {C.~W.}\ \bibnamefont {Gardiner}},\ }\bibfield  {title} {\bibinfo {title} {Squeezing of intracavity and traveling-wave light fields produced in parametric amplification},\ }\href {https://doi.org/10.1103/PhysRevA.30.1386} {\bibfield  {journal} {\bibinfo  {journal} {Phys. Rev. A}\ }\textbf {\bibinfo {volume} {30}},\ \bibinfo {pages} {1386} (\bibinfo {year} {1984})}\BibitemShut {NoStop}%
\end{thebibliography}%
\end{document}